\documentclass[useAMS,usenatbib]{mn2e}

\usepackage{graphicx}

\title[THE NATURE OF THE CORE IN 3C~433] {THE NATURE OF THE NEAR-IR CORE SOURCE IN 3C 433}

\author[Ram\'irez et al.]{Edgar A. Ram\'irez$^{1}$\thanks{E-mail: e.ramirez@sheffield.ac.uk}, C. N. Tadhunter$^{1}$, D. Axon $^{2}$, D. Batcheldor $^{2}$,\newauthor S. Young $^{2}$, C. Packham $^{3}$ and W. B. Sparks $^{4}$\\
$^{1}$Department of Physics and Astronomy, University of Sheffield, Sheffield S3 7RH, UK\\
$^{2}$Department of Physics, Rochester Institute of Technology, 85 Lomb Memorial Drive, Rochester, NY 14623\\
$^{3}$Department of Astronomy, University of Florida, 211 Bryant Science Center, P.O. Box 112055, Gainesville, FL 32611-2055\\
$^{4}$Space Telescope Science Institute, 3700 San Martin Drive, Baltimore, MD 21218}
\begin{document}

\date{Accepted 2009 July 17. Received 2009 July 17; in original form 2009 April 18}

\pagerange{\pageref{firstpage}--\pageref{lastpage}} \pubyear{2009}

\maketitle

\label{firstpage}

\begin{abstract}

We report the analysis of near-infrared imaging, polarimetric and spectroscopic observations of the powerful radio galaxy 3C~433 ($z=0.1016$), obtained with the {\em HST} and UKIRT telescopes. The high spatial resolution of {\em HST} allows us to study the near-nuclear regions of the galaxy ($<1$ kpc). In line with previous observations, we find that 3C~433 has an unresolved core source that is detected in all near-IR  bands, but dominates over the host galaxy emission at $2.05\;\mu$m.

Our analysis reveals: (1) the presence of a dust lane aligned close to perpendicular (PA$=70\pm5\degr$) to the inner radio jet axis (PA$=-12\pm2\degr$); (2) a steep slope to the near-IR SED ($\alpha=5.8\pm0.1$; F$_{\nu}\propto\nu^{-\alpha}$); (3) an apparent lack of broad permitted emission lines at near-IR wavelengths, in particular the absence of a broad  Pa$\alpha$ emission line; and  (4) high intrinsic polarization for the unresolved core nuclear source ($8.6\pm1$ per cent), with an E-vector perpendicular (PA=$83.0\pm 2.3\degr$) to the inner radio jet. Using five independent techniques we determine an extinction to the compact core source in the range $3<$A$_V<67$ mag.

An analysis of the long wavelength SED rules out a synchrotron origin for the high near-IR polarization of the compact core source. Therefore, scattering and dichroic extinction are plausible polarizing mechanisms, although in both of these cases the broad permitted lines from the AGN are required to have a width $>10^4$km s$^{-1}$ (FWHM) to escape detection in our near-IR spectrum. Dichroic extinction is the most likely polarization mechanism because it is consistent with the various available extinction estimates. In this case, a highly ordered, coherent toroidal magnetic field must be present in the obscuring structure close to the nucleus. 

\end{abstract}


\section{INTRODUCTION}

In the unified schemes for radio-loud Active Galactic Nuclei (AGNs),  radio galaxies and quasars are deemed to be the same objects viewed from different lines of sight (Barthel 1989). Some of the best evidence to support such orientation-based unified schemes is provided by spectropolarimetry observations of radio galaxies that show broad permitted lines in the polarized spectrum, characteristic of scatter quasar light (Ogle et al. 1997, Cohen et al. 1999).

A key element of the standard  unified schemes is a central obscuring torus, with an axis parallel to the radio axis. This torus obscures structures close to the nucleus, such as the broad line region (BLR: $r<1$~pc, FWHM$\approx10^3-10^4$ km s$^{-1}$) and the main optical/UV continuum generating region in the accretion disk. However, despite its status as a fundamental element of the unified scheme, large uncertainties remain about the structure of the torus and the degree of extinction it induces.

Apart from helping to clarify the apparent differences among radio-loud AGNs, the circum-nuclear  dust and its spatial distribution are directly affected by phenomena occurring close to the nucleus, such as circum-nuclear outflows. Hence, the study of the distribution and properties of the dust is important for understanding the impact of the nuclear activity on the host galaxies.

Because AGNs are embedded in dusty regions and surrounded by a dust tori, observations at near-infrared (near-IR) wavelengths are crucial, due to the relatively low extinction suffered at such wavelengths. The advantages of observing radio galaxies at near-IR wavelengths have been demonstrated by  imaging observations of Cygnus A (Djorgovski et al. 1991; Tadhunter et al. 1999) and PKS 0634-205 (Simpson, Ward \& Wilson 1995), which show a progression from resolved galaxy-like structures at short near-IR wavelengths, to unresolved point sources at longer wavelengths. These results have been interpreted in terms of the  quasars shining through the obscuring dust torus. In this case, it should be possible to directly detect the BLR associated with the AGNs at near-IR wavelengths. Indeed, previous near-IR spectroscopic observations have shown that some radio galaxies originally classified as narrow line radio galaxies (NLRG) present a  broad Pa$\alpha$ line in the $K$-band (Hill, Goodrich \& DePoy 1996; Holt et al. 2006). This provides evidence for a hidden broad emission line region and gives support to the orientation-based unified schemes. 

3C~433 ($z=0.1016$; Schmidt 1965) is a powerful radio galaxy that has been classified as an NLRG at optical wavelengths (Koski 1978), and  shows signs of a young stellar population in its optical continuum emission (Wills et al. 2002, Holt et al. 2007). With a radio structure comprising jet and knot structures  close to the nucleus, as well as hot spots and distorted radio lobes on larger scales (van Breugel et al. 1983), its Fanaroff and Riley (FR) classification is uncertain, but it is more commonly identified as an FRII source. The existing optical  Hubble Space Telescope {\em HST} images taken with the F702W filter show a complex central structure attributed to dust obscuration and possible star forming regions (de Koff et al. 1996), but no sign of an unresolved optical nucleus (Chiaberge, Capetti \& Celotti 1999). In contrast, 3C~433 has an unusually bright compact nuclear source at near-IR wavelengths (Madrid et al. 2006), as well as exceptionally red near-IR colours (Lilly,  Longair \& Miller 1985). At mid-IR wavelengths, Spitzer observations of 3C~433 indicate the existence of a hidden quasar obscured by dust (Ogle et al. 2006). Therefore, this object provides an excellent opportunity to investigate the nature of compact near-IR core sources in AGNs.

In this article we analyse photometric and polarimetric imaging observations  of the nuclear region of 3C~433 obtained with the Near Infrared Camera and Multi Object Spectrograph (NICMOS) on the {\em HST}. We also present near-IR spectra taken with the United Kingdom Imaging Spectrometer (UIST) Integral Field Unit on the  United Kingdom Infrared Telescope (UKIRT). These data allow us to examine the nature of the nuclear point source using a range of techniques: photometry, spectroscopy, and polarimetry. We analyse the results of these techniques together, and discuss them in the context of the orientation-based unified schemes.
 
In \S 2 we describe the observations, as well as the data reduction procedures. We present the analysis in \S 3, and in \S 4 we discuss the results in terms of the nature of the circum-nuclear structures. In \S 5 we present the conclusions. 

Throughout this paper we assume the following cosmological parameters: $H_0=73$ km s$^{-1}$ Mpc$^{-1}$, $\Omega_M=0.3$, and $\Omega_{\Lambda}=0.7$.

\section[]{OBSERVATIONS AND DATA REDUCTION}

\subsection{{\em HST} IMAGES}

The {\em HST} images were obtained during Cycle 13 on 2005 August 19 (GO~10410, PI: C. N. Tadhunter), using the NICMOS1 camera (NIC1) with the  F110W, F145M, F170M filters (central wavelengths: $1.025$, $1.45$, and $1.7$ $\mu$m respectively), and with the NICMOS2 (NIC2) camera using the  POL-L filters centred  at $2.05\;\mu$m. In the latter case three polarization filter images (POL0L, POL120L and POL240L), whose principal axes of transmission are separated by $120\degr$, were obtained to derive polarimetric information; these were also co-added to form a deep $2.05 \;\mu$m image. 

The observations were coordinated in a parallel way: observations of 3C~433 in NIC1 were made when NIC2 chopped to the sky. The standard X-STRIP-DITHER-CHOP pattern (Barker et al. 2007) was used for the observations in order to facilitate accurate subtraction of the sky emission, with a total of 8 positions (4 sky, 4 object), a chop throw of $31.5$ arcsec (in the Y-direction), and a spacing between the dithering positions of $0.9$ arcsec (in the X-direction). The total on-source integration time were 768 seconds at $1.025$, $1.45$, and $1.7$ $\mu$m, and 624 seconds at $2.05\;\mu$m.

The {\em HST} data were reduced using the standard NICMOS calibration
software {\bf calnica} (MacLaughlin \&  Wiklind  2007). To obtain the polarization degree and position angle using the NIC2 POL-L observations, the Stokes
parameters of the polarimetric data were calculated and analysed
according to the prescription of Sparks \& Axon (1999), as follows. 
First, the mean and standard deviation of the intensity of each polarizer image were obtained through concentric apertures, centred in the galaxy's nucleus, using the IRAF task {\bf phot}. Then, using an IDL routine, we obtained the Stokes parameters ($I,\;U,\;Q$), by solving the matrix of the linear system for three
polarizers. The transmission coefficients for the polarizers POL-L were
derived from data taken with the post-NICMOS Cooling System (Batcheldor et al. 2006). 
Finally, we derived our estimates of the
polarized flux percentage ($100\%\times\sqrt{Q^{2} + U^{2}}/I$) and
the position angles ($\frac{1}{2}{\rm tan}^{-1}(U/Q)$).

\subsection{UKIRT SPECTRA}

In order to search for broad hydrogen emission lines associated with the hidden AGN, we have also analysed spectroscopic data taken with the UIST Integral Field Unit (IFU) imaging spectrometer on UKIRT. 

The observations were obtained with the HK grism on 2003 September 16.  The air mass ranged from $1.089$ at the beginning of the observations, to  $1.661$ at the end. The bandpass extended from $1.41 \;\mu$m to $2.53\;\mu$m, and a mean velocity resolution of $\approx160$ km s$^{-1}$ was achieved. The data were taken in the standard ABBA pattern, with the telescope nodding between the galaxy and sky positions.

The UKIRT data were reduced in the standard way using IRAF. First we combined the galaxy and sky frames separately, using a median filter to remove cosmic rays; then we subtracted sky from galaxy frames to correct for the sky emission, bias and dark current. Residual hot pixels were removed using the {\bf fixpix} task, and the spectra were wavelength calibrated using argon arc frames.  The flux calibration was achieved by dividing by a wavelength calibrated spectrum of the type A0V atmospheric standard star HD34317, observed close in time to the 3C~433 observations; the air mass ranged from $1.47$ to $1.45$. Finally, the data were multiplied by the spectrum of a black body of temperature $9,480$ K to produce a flux and wavelength calibrated spectrum. The spectroscopic analysis presented below focusses on the combined spectrum for a $2\times1$ arcsec aperture centred on the nucleus.

\section[]{ANALYSIS}

\subsection{{\em HST} IMAGING}

\begin{figure*}
  \vspace*{174pt}
  \center\includegraphics[width=5cm, angle=90]{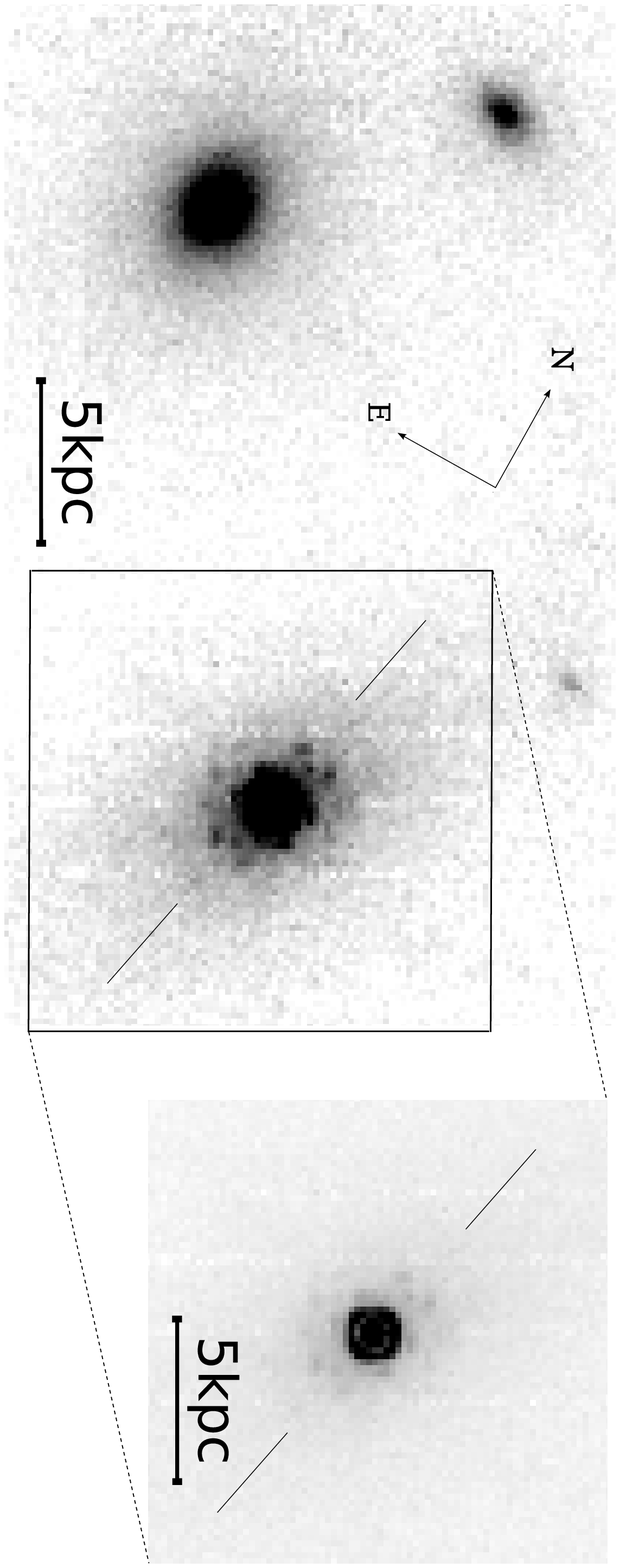}
  \caption{Near-IR images of 3C~433 at $2.05\;\mu$m. The image on the left shows a $13.30\times 7.7$ arcsec field with 3C~433 inside the square; the north and northeast companion galaxies lie to the left side, whereas in the  upper right side lies a patch of emission. The image on the right displays  3C~433 with a different contrast to show the point source quality of the nuclear region. The radio jet orientation is indicated by the pair of lines.}
\label{field}
\end{figure*}

At radio wavelengths 3C~433 is dominated by the extended radio jet and lobes, whereas at optical wavelengths it is dominated by the stellar light of the host galaxy, albeit with a disturbed morphology (de Koff et al. 1996) and signs of recent star formation (Holt et al. 2007). In this work, on the other hand, we find that at near-IR wavelengths, 3C~433 has a bright unresolved point-like nucleus that clearly dominates  over the host galaxy emission (see Fig. \ref{field}). At $2.05 \;\mu$m, the nuclear source shows distinctly the prominent first Airy ring, and also some signs of the second Airy ring. This detection of an unresolved source makes it clear that, using near-IR observations, we are able to penetrate the circum-nuclear dust material. In addition, our  $2.05\;\mu$m POL-L images also detect the nearby companion galaxies $7.8$ arcsec to the North-East, and $9.5$ arcsec to the North of 3C~433 (see also van Breugel et al. 1983), as well the fainter patch of emission $4.1$ arcsec to the North-West of 3C~433, detected before by de Koff et al. (1996).

To estimate the flux of the galaxy core at each wavelength, a trial and error subtraction of PSFs generated by the Tiny Tim software (Krist \& Hook 2004) was performed. The artificial PSFs were re-centred and scaled by different factors to obtain the best subtraction. We consider that the best factor is obtained when, after PSF subtraction, the residual PSF structure is minimised. The fluxes of the scaled PSFs were measured using the GAIA program in the STARLINK software library. The core fluxes derived from this procedure  are listed in Table \ref{flux}.

\begin{table}
  \caption{Core fluxes for 3C~433 measured from {\em HST}/NICMOS images}

  \begin{tabular}{c c c}
    \hline
    Filter & Central wavelength & Flux density (Jy) \\
     &of the bandpass ($\mu$m) & \\
    \hline
    F110W & 1.025 & $(2.8 \pm 0.7) \times10^{-5}$ \\ 
    F145M & 1.45 & $(1.5 \pm 0.1) \times10^{-4}$ \\ 
    F170M & 1.7 & $(4.8 \pm 0.3) \times10^{-4}$ \\ 
    POL-L & 2.05 & $(1.6 \pm 0.2) \times10^{-3}$ \\ 
    \hline
    \end{tabular}\label{flux}
\end{table}

\begin{figure*}
  \vspace*{174pt}
  \center\includegraphics[width=5cm, angle=90]{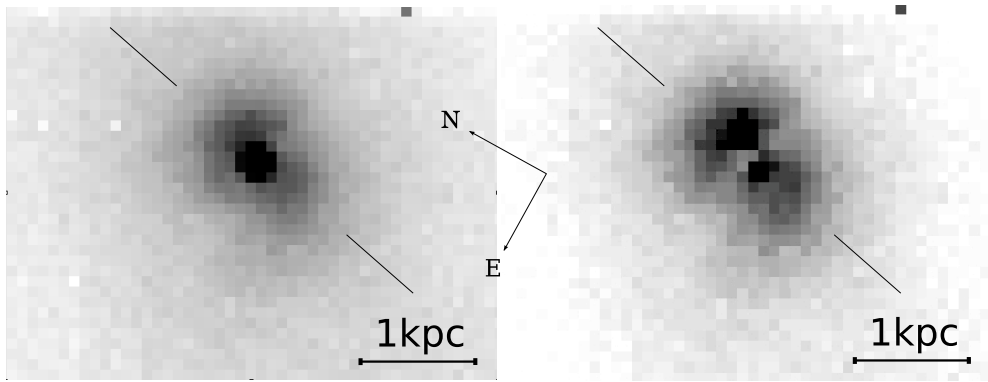}
  \caption{Images of the central region  of 3C~433 at $1.025\;\mu$m. Left: before PSF subtraction, right: after PSF subtraction. The pair of lines represents orientation of radio jets. A dust lane can be seen in both images, but is clearer in the right hand image.}
\label{dusties}
\end{figure*}

While the $2.05\;\mu$m image is dominated by an unresolved point source, at shorter wavelengths ($1.025 \;\mu$m), using the resolution of {\em HST} (0.1 arcsec) it is possible to detect for the first time a kpc-scale dust lane. In Fig. \ref{dusties} we compare the  $1.025 \;\mu$m image before and after PSF subtraction. The position angle of the dust lane is PA$=70\pm5\degr$. For comparison, the inner jet axis defined by the compact core components C1 and C2, detected in the $8.465$ GHz map of Black et al. (1992), has PA$=-12\pm2\degr$, whereas the outer jet axis, defined by components S1 and N1 from Black et al. (1992), has PA$=-14.0\pm 0.5\degr$. Therefore, the position angle of the dust lane is offset by $82\pm 5$ degrees relative to the inner jet, and $84\pm5$ degrees with respect to the outer jet.

\subsubsection{POLARIMETRY}

Using an aperture of $0.9$ arcsec diameter centred on the nuclear source (this contains the 1st Airy ring), we derive a relatively high linear  polarization of P$_{2.05}=7.0\pm0.2$ per cent. The same analysis applied to the nucleus of the brighter northeast companion galaxy gives P$_{2.05}=0.5\pm1$ per cent. This demonstrates that the high near-IR  polarization measured in the nucleus of 3C~433 is not due to systematic or calibration errors. Note that, due to the dominance of the PSF of the nuclear point source, we have not been able to measure significant spatially resolved polarization in the halo of the galaxy. 

The degree of polarization is underestimated because unpolarized starlight from the host galaxy dilutes the intrinsic AGN polarization. We have estimated the host galaxy starlight contribution in the following two ways. First,  we measured the starlight  contribution  from the  $1.025 \;\mu$m image\footnote[1]{Since the AGN point source contributes least at the shorter wavelengths, the $1.025 \;\mu$m can potentially provide the most accurate measure of the starlight contribution, although additional uncertainties arise because of the need to extrapolate from $1.025 \;\mu$m to $2.05\;\mu$m.} following subtraction of the best fitting PSF for the point source, then  extrapolated to $2.05\;\mu$m using the typical colours for radio galaxies hosts  from  Lilly,  Longair \& Miller (1985). This yields a starlight contribution of $14$ per cent to the total flux at $2.05\;\mu$m. Second, we have measured the starlight contribution  directly from the  PSF-subtracted $2.05\;\mu$m image, giving a  starlight contribution of $23$ per cent. Both  methods  used a $0.9$ arcsec diameter aperture. Assuming that the starlight is itself unpolarized and contributes $14$-$23$ per cent of the total light in the polarimetric aperture, we therefore estimate an intrinsic polarization of P$^{int}_{2.05}=8.6\pm1$ per cent for the point source.

The PA of the E-vector for the nuclear point source is $83\pm2.3\degr$,  offset by $95\pm3\degr$ relative to the inner radio jet and $97\pm2.3\degr$ relative to the outer radio jet. That is, the radio axis and $2.05\;\mu$m-band polarization vector are close to perpendicular within the estimated uncertainty.

\subsubsection{SED AND EXTINCTION}

\begin{figure*}
  \vspace*{174pt}
  \center\includegraphics[width=11cm]{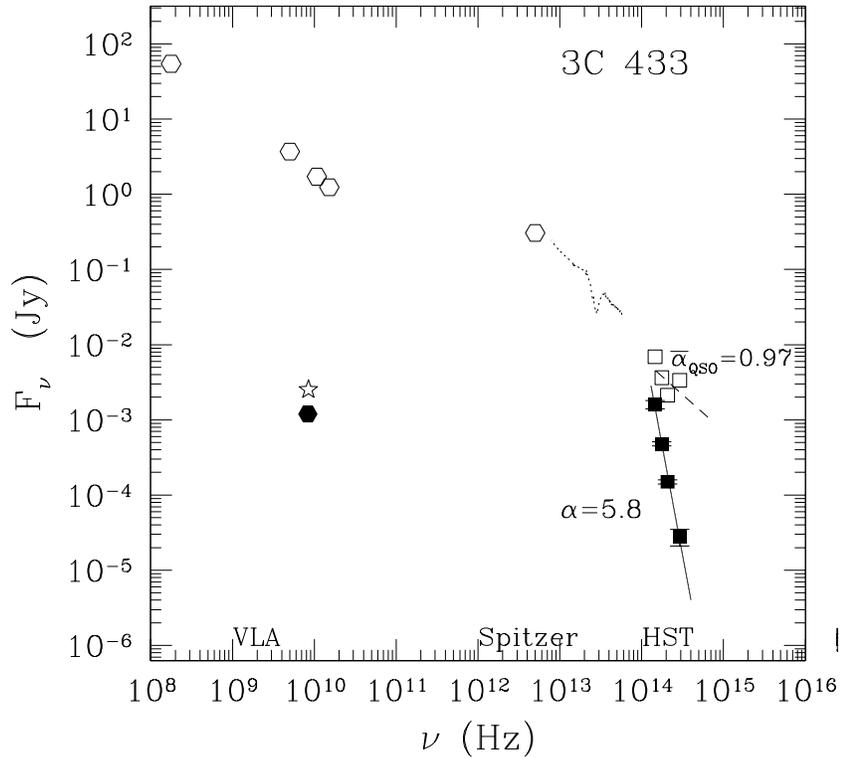}
  \caption{The spectral energy distribution (SED) of 3C~433. The {\em HST}/NICMOS fluxes of the unresolved core source are represented by filled  squares; the solid straight line is the best power law fit ($\alpha=5.8\pm0.1$). The dashed line is the mean de-reddened power-law for quasars ($\overline{\alpha}_{QSO}=0.97$: Simpson \& Rawlings 2000). The open squares are the reddening corrected  {\em HST}/NICMOS points. The open hexagons are radio-VLA data from NED and the  $60\;\mu$m far-IR flux from Impey \& Gregorini (1993), respectively. The filled hexagon and the star are the radio emission from the southern radio core component of Hardcastle et al. (1998) and our own measurements, respectively. We have measured the VLA $8.465$ GHz flux from the data published in (Black et al. 1992) of the south component of the radio core (C2), which is marginally more compact than the northern core component (Black et al. 1992), and coincident with the Chandra position of the nucleus  (Hardcastle 2008, private communication). We find a $8.45$ GHz flux of $2.53\times10^{-3}$ Jy for the southern knot, the slightly  discrepancy between our radio measurements and that of Hardcastle et al. (1998) is likely to be due to the fact that the Hardcastle et al. (1998) core flux has been corrected for contamination by the underlying radio jet. Finally, the doted line represents Spitzer IRS data (Ogle et al. 2006).} 
\label{SED}
\end{figure*}

The spectral energy distribution (SED) of 3C~433 is shown in Fig. \ref{SED}. Clearly, the flux from the nuclear point source shows a steep decline towards shorter wavelengths in the near-IR. Fitting a power-law to the {\em HST}/NICMOS core fluxes ($F_{\nu}\propto\nu^{-\alpha}$, $F_{\lambda}\propto\lambda^{\alpha-2}$), we obtain a spectral index $\alpha=5.8\pm0.1$. 


In order to derive the degree of reddening of the nuclear point source we have used five different techniques.

(1) Near-IR SED. The de-reddening necessary to make the measured near-IR slope consistent with a quasar spectrum ($-0.67<\alpha<1.62$, with $\overline{\alpha}_{QSO}=0.97$; Simpson \& Rawlings 2000) was estimated. This method gives an extinction A$_{2.05\mu{\rm m}}=1.6\pm^{0.6}_{0.2}$ mag. Extrapolating to optical wavelengths using a typical Galactic extinction law (A$_{2.05\mu{\rm m}}/$A$_V=0.107$; Mathis 1990) and correcting for Galactic extinction (A$_V=0.68$ mag; Schlegel, Finkbeiner \& Davis 1998), A$_V=14\pm^{6}_{2}$ mag is obtained.

(2) X-ray luminosity. We estimated the $1.0\;\mu$m luminosity using the X-ray to near-IR correlation in Kriss (1988) and the nuclear luminosity of  3C~433 at 2keV  determined by  Hardcastle (2008 private communication)  from Chandra data. Next, we find  F$_{1.025\mu{\rm m}}$, taking the luminosity distance of 444 Mpc (from NED). By comparing this flux to our observed $1.025 \;\mu$m flux we derive an extinction of A$_{1.025\mu{\rm m}}=6.2\pm0.4$ mag. Finally,  A$_V=17.8\pm1.1$ mag is obtained using the extinction law of Mathis (1990).

(3) X-ray column. The atomic hydrogen column density derived from Chandra data is N$_{H}=9\pm 1\times 10^{22}$ cm$ ^{-2}$ (Hardcastle 2008, private communication). Using the standard Galactic ratio A$_V$/N$_H=5.3\times10^{-22}$ mag cm$^2$ (Bohlin, Savage \& Drake 1978), A$_V= 48\pm5$ mag is obtained. However, given that AGNs generally have dust to  gas ratios that are significantly lower than Galactic (Maiolino et al. 2001), this is likely to represent an upper limit. We have used the results from Maiolino et al. (2001) shown in his Fig. 1, to calculate the upper and lower limits of true dust extinction. Referring to Fig 1. of Maiolino et al. (2001), we calculated the mean $E_{B-V}/N_H$ ratio of the seven upper AGNs below the Galactic Standard, and the mean of the ratio of the lower points, both using the ADS's data extractor DEXTER (Demleitner et al. 2001). With these two mean ratios, we find that the true dust extinction is likely to fall in the range $3<$A$_V<17$ mag.

(4) Near-IR polarization. Under the assumption that the measured intrinsic near-IR polarization is due to dichroic extinction by near-nuclear dust structures, the correlation between the observed linear polarization in the $K$-band and the optical extinction (Jones 1989) yields $15<$A$_V<67$ mag, i.e., consistent with our previous estimates of A$_V$. 

(5) Silicate absorption. The Spitzer IR spectrum shows a deep silicate absorption line at $9.7\mu$m, with an optical depth of $\tau_{9.7}=0.71\pm 0.07$ (Ogle et al. 2006). Using the empirically derived relationship between optical extinction and silicate absorption strength (A$_{V}/\tau_{9.7}=18.5\pm1$: Whittet 1987), we find A$_{V}=13.1\pm3.3$. 

The various extinction estimates are compared in Table \ref{A_V}. Overall, given the uncertainties inherent in all the techniques, there is a remarkable degree of agreement.  The similarity between the extinction estimated using the dichroic method and the other extinction estimates support a dichroic origin of the polarization (see the discussion section).  However, scattering processes can also make an important contribution to the polarization.

\begin{center}
\begin{table}

  \caption{Extinction estimates for the core source in 3C~433, based on various methods . The results are compared with those obtained for Cygnus~A using the same methods.}
  \begin{tabular}{c l l l}
    \hline
    No. method & Method & \multicolumn{2}{c}{A$_V$ (mag) }\\
    & & 3C~433& Cygnus A\\
     \hline
    1 & near-IR SED  & $14\pm^{6}_{2}$&$-$\\
    2 & X-ray luminosity & $18\pm1$ & $107$\\
    3 & X-ray column & $3-17$ & $11-69$\\
    4 & Near-IR polarization & $15-67$ & $50-324$\\ 
    5 & Silicate absorption & $13\pm3$ & $-$\\ 
    \hline
    \end{tabular}
\label{A_V}
\end{table}
\end{center}

\subsection{UKIRT SPECTROSCOPY}

\begin{figure*}
  \vspace*{174pt}
  \center\includegraphics[width=12cm]{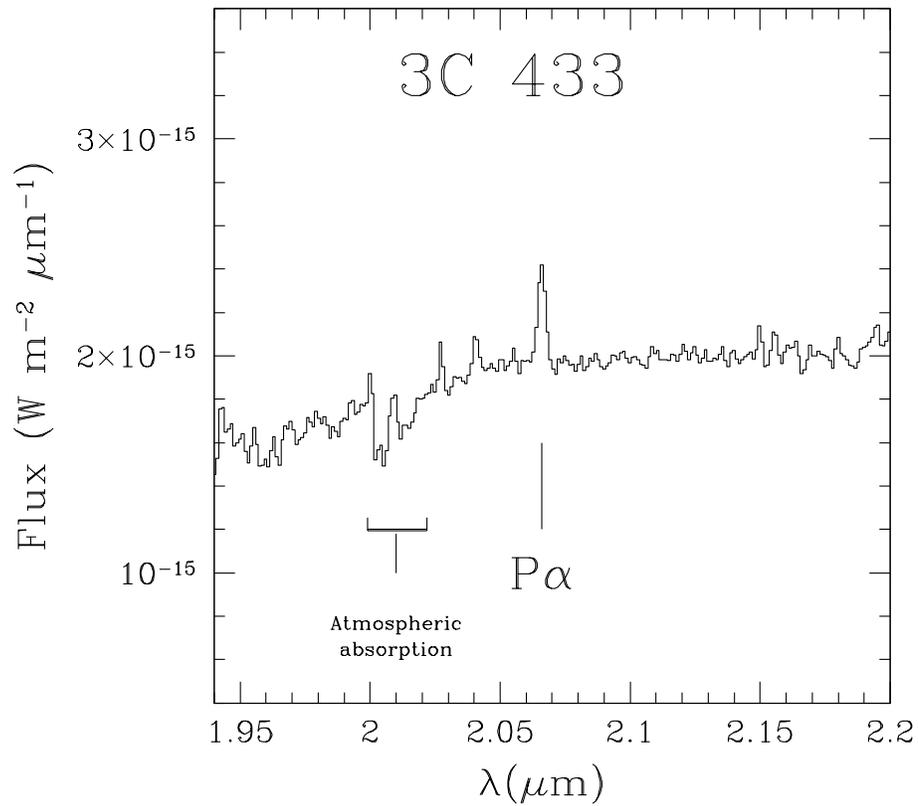}
  \caption{Near-IR UKIRT/UIST spectrum of the radio galaxy 3C~433 from $1.94$ to $2.2\;\mu$m. The position of the narrow Pa$\alpha$ line is marked. Residual atmospheric absorption features are visible around $2.0\;\mu$m that were not completely removed by division by the telluric standard star spectrum. Note the lack of any clear evidence for a broad Pa$\alpha$ feature.}
\label{spec}
\end{figure*}

Despite the strength of the near-IR continuum from the unresolved AGN, the UKIRT spectrum of the nucleus shows no clear evidence for a strong broad Pa$\alpha$ line.  After removing telluric absorption features, a narrow  Pa$\alpha$ ($\lambda_0=1.8756\; \mu$m) emission line is present at the expected redshifted wavelength ($\approx 2.07\;\mu$m; see Fig. \ref{spec}).

This lack of a broad line is unexpected, and stands out for the following reasons. Our near-IR results suggest that we are able to look through the dust torus directly into the AGN. A strong broad Pa$\alpha$ line should thus be present. Moreover, a prominent broad Pa$\alpha$ is detected in the spectra of other powerful radio galaxies with obscured quasar nuclei (e.g., PKS 1549-79; Holt et al. 2006).

\begin{figure*}
  \vspace*{174pt}
  \center\includegraphics[width=15cm]{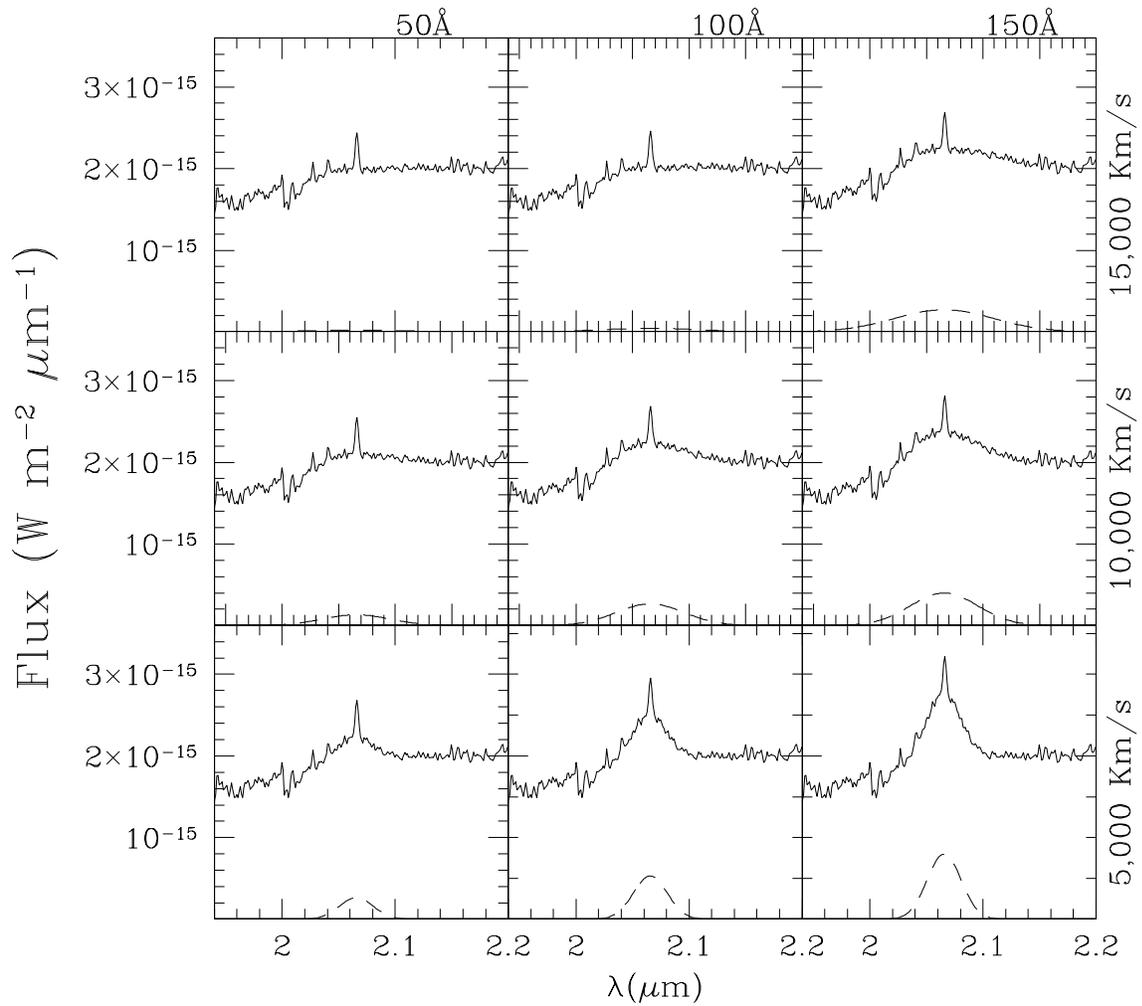}
  \caption{The spectral energy distribution of 3C~433 (solid lines) with added Gaussian profiles (dashed lines) representing broad Pa$\alpha$ features with different equivalent widths (columns) and FWHM (rows). For a typical quasar Pa$\alpha$ equivalent width (EW$= 100\;\rm\AA$) it is difficult to detect the broad Pa$\alpha$ line with these data if the lines has FWHM$\geq10,000$ km s$^{-1}$.}
 

\label{gaussians}
\end{figure*}

Although we do not detect a broad Pa$\alpha$ line directly in our spectra, we can estimate the ranges of Pa$\alpha$ equivalent width (EW) and FWHM that are consistent with the non-detection. This was done by adding broad Gaussian lines of various velocity widths (FWHM$=5,000$, $10,000$ and $15,000$ km s$^{-1}$) and equivalent widths ($50$, $100$ and $150\;\rm \AA$) to our spectrum at the expected wavelength of Pa$\alpha$, and then examining whether the line could be clearly detected above the continuum upon visual inspection of the plots (see Fig. \ref{gaussians}). Note that the equivalent widths of the broad Pa$\alpha$ line in optically obscured and unobscured quasar nuclei generally fall in the range  $50<$ EW$_{\rm Pa\alpha}<170 \;\rm\AA$ (Rudy \& Tokunaga 1982, Holt et al. 2006, Landt et al. 2008).

On the basis of Fig. \ref{gaussians}, the non-detection of broad Pa${\alpha}$ is favoured by a large FWHM and a small EW. If the AGN in 3C~433 has a broad Pa$\alpha$ line with EW typical of the AGNs (median EW$\approx92\;\rm\AA$: Landt et al. 2008), its Pa$\alpha$ must have a velocity width $>10,000$ km s$^{-1}$ to escape detection. Such large velocity widths, while not typical of AGNs in general, are commonly measured in broad line radio-loud AGNs  in the local Universe, which have radio and emission line luminosities similar to 3C~433 (e.g. the BLRG in Osterbrock, Koski \& Phillips 1976, Buttiglione et al. 2009). Therefore, the simplest explanation for the non-detected broad Pa$\alpha$ in 3C~433 is that it contains a typical BLRG nucleus with FWHM$>10,000$ km s$^{-1}$.

\section{DISCUSSION}

We have found that 3C~433 shows several interesting characteristics at near-IR wavelengths. In this section we summarize and discuss our results. 

\begin{itemize}

\item An unresolved point-like nucleus is detected at all near-IR wavelengths. This source dominates  over the starlight from the host galaxy in the $2.05\;\mu$m-band.  The strength of the nuclear point source suggests that we are looking directly at the AGN through the circum-nuclear dust.

\item It is plausible that the kpc-scale dust lane that falls close to perpendicular to the inner radio jet ($\Delta\theta=82\pm5\degr$) forms the outer part of a torus structure responsible for the nuclear obscuration\footnote[2]{Hereafter we refer to the ``circum-nuclear obscuring region'' as the  ``dust torus''. Because there may exist dust on a continuous  range of scales between the inner torus and outer kpc-dust lane, the division between torus and dust lane is largely semantic.}.

\item The extinction obtained using the different methods is consistent with a relatively modest obscuration of the nuclear point source ($3<$A$_V<67$ mag), compared with the archetypical powerful radio galaxy Cygnus~A (Tadhunter et al. 1999).

\item We have found that the spectrum of 3C~433 lacks a strong  broad Pa$\alpha$ emission line. This is unexpected, because the IR images suggest that we are looking directly at the AGN. However, it is plausible that the non-detection is due to the fact that 3C~433 has a Pa$\alpha$ line with a relatively large width (FWHM$>\;10,000$ km s$^{-1}$; typical of BLRG), which makes it difficult to detect in our spectra.

\item The nucleus is highly polarized (P$^{int}_{2.05\;\mu {\rm m}}=8.6\pm1$ per cent); and the orientation of the polarization E-vector at near-IR wavelengths is close to  perpendicular to the inner radio jet axis ($\Delta\theta=95\pm3\degr$). 

The polarization of the near-IR light could be produced by three mechanisms: (1) non-thermal synchrotron radiation; (2) scattering of the light due to dust or electrons; or  (3) dichroic extinction.

At first sight, it might be thought that the lack of a broad Pa$\alpha$ line and the relatively high degree of polarization can be explained in terms of strong non-thermal synchrotron continuum  emission swamping the emission lines at near-IR wavelengths. Indeed, Capetti et al. (2007) have proposed a synchrotron origin for the optical polarization measured in the unresolved optical cores of a sample of 9 nearby FRI radio galaxies. Such interpretation is supported by the close correlations between the X-ray, optical and radio properties of the FRI cores (e.g. Balmaverde, Capetti \& Grandi 2006). However, in 3C~433 a major contribution to the near-IR continuum emission from a synchrotron mechanism is unlikely for the following reasons. The SEDs produced by synchrotron radiation from flat spectrum core radio sources generally show significant declines -- by orders of magnitude -- between the radio and the near-IR wavelengths (e.g. Dicken et al. 2008). A clear example is the quasar 3C~273, for which the flat spectrum radio core is two orders of magnitude brighter than its near-IR core flux. Moreover, the flat spectrum radio cores in lower-luminosity FRI radio galaxies also have relatively steep spectra between radio and optical wavelengths ($\alpha\approx 0.6-1.0$, Balmaverde, Capetti \& Grandi 2006)\footnote[3]{Note that, in the case of the FRI sources, there is no evidence that the compact cores suffer substantial dust extinction at optical wavelengths (typical A$_V$ between $0.15$ and $3$ mag: Chiaberge et al. 2002), and the X-ray absorption column densities in these sources are also relatively low (Balmaverde, Capetti \& Grandi 2006).}. In contrast, a power-law fit between the radio core flux and the near-IR data for 3C~433 yields a flat or gently rising spectrum (see Fig. \ref{SED}), with no decline between the radio and the near-IR wavelengths. Given that this is  difficult to reconcile with the steep radio-IR SEDs of known synchrotron dominated core sources, it is unlikely that synchrotron radiation substantially dilutes the broad  Pa$\alpha$ feature.

Regarding scattering, we expect the scattered light to be polarized perpendicular to the radius vector from the illuminating source. We have found that the PA of the E-vector is close to parallel to the dust lane (offset$=13.0\pm5.5\degr$) and perpendicular to the inner radio axis (offset$=95\pm3\degr$). Therefore, we could be seeing  the light scattered by the inner face of the torus on the far side of the nucleus, or polar scattering  by material above the torus. Potentially, scattered light could provide an alternative to explain the non-detection of broad Pa$\alpha$ line, since,  if the scattering is by electrons and they are sufficiently hot ($>10^{6}$K), then a very broad Pa$\alpha$ line (FWHM$>10,000$ km s$^{-1}$) would not be detected in our spectroscopic data if it has an equivalent width typical of quasars.

Dichroic extinction is the selective attenuation of light passing through a medium with dust grains that are aligned by a magnetic field. It has been found that the efficiency of this mechanism is low: large quantities of dust are required to produce a reasonable degree of polarized light (Jones 1989). In the case of 3C~433, the similarity between the extinction required to give the measured polarization  at $2.05\;\mu$m and the other extinction estimates (see Table \ref{A_V}) provides strong evidence for a dichroic origin of the polarization. If this is the case, since the E-vector of the polarized light is aligned perpendicular to the radio axis, the dust grains in the torus must be aligned perpendicular to the torus plane and, therefore, the  magnetic field must be  parallel to the dust torus (i.e., the magnetic field is toroidal). This scheme requires the presence of elongated dust grains and the existence of a highly coherent magnetic fields in the torus. Indeed, for the dichroic mechanism to work the polarization efficiency is required to be close to the upper limit of that measured for stars in our galaxy (Jones 1989). This can be explained in terms of the fact that the circum-nuclear torus is likely to be more compact and coherent structure than the large-scale disk of the Milky Way.

It is also interesting to compare our results for 3C~433 with those obtained for Cygnus~A, the only other powerful radio source to have its near-IR core source investigated in such detail.  We have re-calculated the optical extinction in Cygnus~A using the various methods applied in 3C~433, and tabulated the results in Table \ref{A_V}. By method (2) we obtain  A$_V=107$ mag, given the nuclear X-ray luminosity in Ueno et al. (1994). This optical extinction is slightly higher than the reported in Tadhunter et al. (1999; $93$ mag), because the X-ray luminosity given by Ueno et al. (1994) is higher than the one taken in Tadhunter et al. (1999).  By method (3) we estimate A$_V=198$ mag, given the column density  of N$_{\rm H}=3.75\times10^{23}$ cm$^{-2}$ (Ueno et al. 1994), and by applying the arguments of Maiolino et al. (2001) we get A$_V=11-69$ mag. The intrinsic  near-IR polarization measured in the unresolved core of Cygnus~A is P$_K=28$ per cent, with an offset of $96\degr$ between the radio jet and the E-vector (Tadhunter et al. 2000). Under the assumptions of method (4), we obtain  $50<$A$_V<324$ mag. 

Its notable that the ratio of IR polarization to visual extinction estimated using methods (2) and (3) covers a similar range for Cygnus A ($0.26<$P$_K/$A$_V<2.5$) and for 3C~433 ($0.47<$P$^{int}_K/$A$_V<2.9$). The fact that there is evidence for higher  extinction in the case of Cygnus~A helps to explain why its near-IR core source is relatively fainter than that detected in 3C~433.

Additional polarimetry observations at other near- and mid-IR wavelengths could help to distinguish among the three polarization mechanism mentioned above. In particular, comparison of our polarimetric measurements at $2.05\;\mu$m with those at both longer and shorter wavelengths could be used to distinguish between the scattering and dichroic polarization mechanisms. In the case of electron or optically thick dust scattering we expect the polarization to show little variations across the near- and mid-IR bands, whereas the dichroic mechanism will produce a significant decline between $J$ and $L$ bands, following the Serkowski law (Serkowski 1973).

\end{itemize}

\section{CONCLUSIONS}

We have used {\em HST} and UKIRT observations to study 3C~433 at near-IR wavelengths, and have found several interesting features. The near-IR observations suggest that we are viewing the AGN directly through an obscuring dust structure perpendicular to the radio jets, or indirectly via scattering by a region close to the nucleus. On the other hand, the near-IR spectrum of the source shows a lack of broad lines. However, the estimated level of extinction ($3<$A$_V<67$ mag) and high intrinsic  polarization of the nucleus ($8.6\pm1$ per cent) support the idea that the high  polarization  may be due to dichroic extinction by a magnetised torus structure in the nuclear region of the galaxy.

It is now important to extend such studies to larger samples, in order to determine  whether the polarization and extinction properties of 3C~433 are typical of the general population of powerful radio galaxies.

\section*{ACKNOWLEDGEMENTS}
EAR acknowledges financial support from the Mexican Council of Science and Technology (CONACyT). Thanks to Rosa A. Gonz\'alez-Lopezlira for her comments, and for an IRAF routine to perform the PSF subtraction; thanks also to Martin Hardcastle for allowing us the use of his X-ray results data. Also we would like to thank the  constructive comments of the anonymous referee. Based on observations collected with NASA/ESA  Hubble Space Telescope, operated by the Association of Universities for Research in Astronomy under contract with NASA, NAS5-26555, and with the United Kingdom Infrared Telescope, operated by the Joint Astronomy Centre on behalf of the Science and Technology Facilities Council of the U.K. This research makes use of the NSA/IPAC extragalactic database (NED).

\bsp

\label{lastpage}

\end{document}